\documentclass[twoside,fleqn]{article}
\usepackage{espcrc2,epsf}

\title{
\vspace{-3cm} {\normalsize \hspace{12cm} HUPD-9715}\\
\vspace{2cm}
NRQCD Results on Form Factors\thanks{Talk presented at
the International workshop ``Lattice QCD on Parallel Computers''
,March 1997, Tsukuba}} 

\author{S. Hashimoto\address{
    High Energy Accelerator Research Organization (KEK), 
    Tsukuba 305, JAPAN},
  K. Ishikawa\address{
    Department of Physics, Hiroshima University,
    1-3-1 Kagamiyama, Higashi-Hiroshima 739, JAPAN},
  H. Matsufuru$^{\rm b}$,
  T. Onogi$^{\rm b}$ and
  N. Yamada$^{\rm b}$ }

\begin{document}

\begin{abstract}
  We report results on $f_B$ and semi-leptonic $B$ decay form 
  factors using NRQCD. 
  We investigate $1/M$ scaling behavior of decay amplitudes.  
  For $f_B$ Effect of higher order relativistic correction
  terms are also studied.
\end{abstract}

\maketitle

\section{Introduction}
Weak matrix elements of $B$ meson such as 
$f_B$, $B_B$, and $B \rightarrow \pi(\rho)l\nu$ form factors 
are important quantities for the determination of
Cabbibo-Kobayashi-Maskawa matrix elements. 
However simulating the $b$-quark with high precision is
still a challenge in Lattice QCD, since the $b$-quark mass
in the lattice unit is large, $a m_b \sim$ 2--3, even
in recent lattice calculations.
One approach to deal with the heavy quark is to extrapolate
the matrix elements for heavy-light meson around the charm
quark mass region to the $b$-quark mass assuming $1/m$
scaling. 
It is, however, rather difficult to control the systematic
uncertainty in this approach, since the systematic error tends
to become larger as increasing $am_Q$.
An alternative approach is to use an effective
nonrelativistic action.  
In Table \ref{tab:action}, we compare various features of
NRQCD\cite{NRQCD}, Fermilab\cite{FNAL}, and ordinary
Wilson/Clover actions as $b$-quark action.   
The first two actions are the effective nonrelativistic
actions. 
Their advantage is that $b$-quark can be directly simulated.

\begin{table}
\caption{Fermion actions for heavy quark.}
\label{tab:action}
\begin{center}
\begin{tabular}{|c|c|c|}
\hline
NRQCD           & Fermilab      & Wilson/Clover\\
\hline
\hline
Direct          & Direct        & Extrapolation\\
Simulation      & Simulation    & from charm\\
\hline
\hline
Error size      & Error size & Error size\\
  $\frac{\alpha_s \Lambda}{m_Q} ,\frac{\Lambda^2}{m_Q^2}$
& $\frac{\alpha_s \Lambda}{m_Q}, \frac{\Lambda^2}{m_Q^2}$       
& $(a m_Q)^2, \alpha_s a m_Q$   \\
\hline
Error remains       & Error $\rightarrow$ 0 & Error $\rightarrow$ 0 \\ 
as $\beta \nearrow$ & as $\beta \nearrow$   & as $\beta \nearrow$ \\
\hline
Easily          &               & \\
Improvable      & Improvable    &  \\
\hline
\end{tabular}
\end{center}
\end{table}

In this report, we present our study of the decay constant
of $B$ meson and $B \rightarrow \pi l\overline{\nu}$
semi-leptonic decay form factors with NRQCD action.
We study the mass dependence of these quantities by
simulating heavy-light mesons over a wide range of the heavy
quark mass.  
In section \ref{sec:decay_const},
we study the 1/m dependence of the heavy-light decay
constant using nonrelativistic action of $O(1/m_Q^2)$. 
The systematic errors due the truncation of higher order 
relativistic correction terms are estimated.
In section \ref{sec:form_factor} we describe the first
computation of the $B \rightarrow \pi l\overline{\nu}$
semi-leptonic decay form factors with NRQCD action of
$O(1/m_Q)$.
Section \ref{sec:discussion} is devoted for discussion and 
future problems.

\section{$B$ meson decay constant $f_B$}
\label{sec:decay_const} 
In the NRQCD approach it is very important to investigate
the size of the systematic error arising from the truncation 
of the action at a certain order of $1/m_{Q}$.
Earlier studies on $f_B$ by Davies et al.\cite{fb1} and
Hashimoto\cite{fb2} and the subsequent work by NRQCD
group\cite{NRfB}, where $b$-quark is simulated with
the NRQCD action including up to $O(1/m_{Q})$ terms, showed
that $1/m_P$ correction of $f_B$ from the static limit is
significantly large. 
Thus the effect of the higher order correction of
$O(1/m_Q^2)$ could be important.
In this section, we compare the $B$ meson decay constant
obtained from the action including $O(1/m)$ terms only and
that including $O(1/m^2)$ terms entirely.


\subsection{NRQCD action and field rotation}
\label{sec:action_fB}

We employ the following NRQCD action
\begin{eqnarray}
  S & = & Q^{\dagger}(t,\mbox{\boldmath $x$})
      \left[ Q(t,\mbox{\boldmath $x$})-
      \left( 1 - \frac{aH_0}{2n} \right)^n \right.
  \nonumber\\  
  & & \times \left( 1 - \frac{a\delta H}{2} \right)
      U_4^{\dag}
      \left( 1 - \frac{a\delta H}{2} \right) 
  \nonumber\\  
  & & \times \left. \left( 1 - \frac{aH_0}{2n} \right)^n 
      Q(t-1,\mbox{\boldmath $x$}) \right]
\end{eqnarray}
where
\begin{eqnarray}
  H_0            & = & - \frac{\Delta^{(2)}}{2m_Q}, \\
  \delta H       & = & \sum_i c_i \delta H^{(i)},   \\
  \delta H^{(1)} & = & - \frac{g}{2m_Q}
     \mbox{\boldmath $\sigma$} \cdot \mbox{\boldmath $B$},\\
  \delta H^{(2)} & = &   \frac{ig}{8m_Q^2}
     ( \mbox{\boldmath $\Delta$} \cdot \mbox{\boldmath $E$}
     - \mbox{\boldmath $E$}\cdot \mbox{\boldmath $\Delta$} ),\\
  \delta H^{(3)} & = & - \frac{g}{8m_Q^2} \sigma \cdot 
     (\mbox{\boldmath $\Delta$}\times\mbox{\boldmath $E$}
     - \mbox{\boldmath $E$} \times \mbox{\boldmath $\Delta$}), \\
  \delta H^{(4)} & = & - \frac{(\Delta^{(2)})^2}{8m_Q^3} ,\\
  \delta H^{(5)} & = &   \frac{a^2 \Delta^{(4)}}{24m_Q},        \\
  \delta H^{(6)} & = & - \frac{a (\Delta^{(2)})^2}{16nm_Q^2},
\end{eqnarray}
where $n$ denotes the stabilization parameter.
The coefficients $c_i$ are unity at tree level and should be
determined by perturbatively matching the action to that in
relativistic QCD in order to include the 1-loop corrections.
$\Delta$ and $\Delta^{(2)}$ denote the symmetric lattice
differentiation in spatial directions and Laplacian
respectively and $\Delta^{(4)} \equiv \sum_i (\Delta^{(2)}_i)^2$.
$\mbox{\boldmath $B$}$ and $\mbox{\boldmath $E$}$ are
generated from the standard clover-leaf field strength.

The original 4-component heavy quark spinor $h$
is decomposed into two 2-component spinors $Q$ and $\chi$
after Foldy-Wouthuysen-Tani (FWT) transformation:
\begin{eqnarray}
h(x) & = & R \left( \begin{array}{c} Q(x) \\ \chi^{\dag}(x) 
                    \end{array} \right),
\end{eqnarray}
where $R$ is an inverse FWT transformation matrix which has
$4\times4$ spin and $3\times3$ color indices.
After discretization, at the tree level $R$ is written as
follows: 
\begin{eqnarray}
  R       & = & \sum_{i}R^{(i)}, \\
  \label{eq:R(1)}
  R^{(1)} & = &  1, \\
  \label{eq:R(2)}
  R^{(2)} & = & - \frac{\mbox{\boldmath $\gamma$} 
    \cdot \mbox{\boldmath $\Delta$} }{2m_Q}, \\
  \label{eq:R(3)}
  R^{(3)} & = & \frac{\mbox{\boldmath $\Delta$}^{(2)}}{8m_Q^2}, \\
  \label{eq:R(4)}
  R^{(4)} & = & \frac{g\mbox{\boldmath $\Sigma$}
      \cdot \mbox{\boldmath $B$}}{8m_Q^2} , \\ 
  \label{eq:R(5)}
  R^{(5)} & = & - \frac{ig\gamma_4 \mbox{\boldmath $\gamma$} 
    \cdot{\mbox{\boldmath $E$} }}{4m_Q^2},   
\end{eqnarray}
where
\begin{eqnarray}
\Sigma^{j} = \left( \begin{array}{cc} \sigma^{j} &     0      \\
            0      & \sigma^{j} \\ \end{array} \right).
\end{eqnarray}
We apply the tadpole improvement\cite{MFimp} to all link
variables in the evolution equation and $R$ by 
rescaling the link variables as $U_{\mu} \rightarrow
U_{\mu}/u_0$. 

We define two sets of
action and current operator \{$\delta H$,$R$\} as follows,
\begin{eqnarray} 
  \mbox{setI}  \equiv \{ \delta H_I,R_I \}
  \; & \mbox{and} & \;
  \mbox{setI$\!$I} \equiv \{ \delta H_{I\!I},R_{I\!I} \},
\end{eqnarray}
where
\begin{eqnarray}
  \delta H_I = \delta H^{(1)} 
  \; & \mbox{and} & \;
  R_I = \sum_{i=1}^{2}R^{(i)}, 
	\label{eq:R1}                \\
  \delta H_{I\!I} = \sum_{i=1}^{6}\delta H^{(i)}
  \; & \mbox{and} & \; 
  R_{I\!I} = \sum_{i=1}^{5}R^{(i)}.
\end{eqnarray}
$\delta H_1$ and $R_1$ include only $O(1/m_Q)$ terms while 
$\delta H_2$ and $R_2$ keep entire $O(1/m_Q^2)$ terms and
the leading relativistic correction to the dispersion
relation, which is an $O(1/m_Q^3)$ term.
The terms improving the discretization errors appearing in
$H_0$ and time evolution are also included.

Using these two sets, we can realize two
levels of accuracy of $O(1/m_Q)$ and $O(1/m_Q^2)$.

\subsection{Simulation methods}
We have computed $f_B$ at $\beta=5.8$ on 120 $16^3\times32$ 
lattices with periodic boundary condition in the spatial 
direction and Dirichlet boundary condition in the temporal
direction. 
The inverse lattice spacing $a^{-1}$ determined from
$m_{\rho}$ is 1.714(63) GeV.
For heavy quark, we use both mean-field improved $O(1/m_Q)$
and $O(1/m_Q^2)$ NRQCD action.
We take six points for the heavy quark mass $a m_Q$ in a
range 0.9--5.0 (1.5--8.5 GeV ).
For light quark, we use Wilson action at $\kappa$= 0.1600,
0.1585, and 0.1570 ($k_{\rm crit}=0.16337$) which
correspond to $m_s$--$2 m_s$. 


\subsection{Results}
\begin{figure}[tb]
  \begin{center}
    \epsfxsize=7.0cm \epsffile{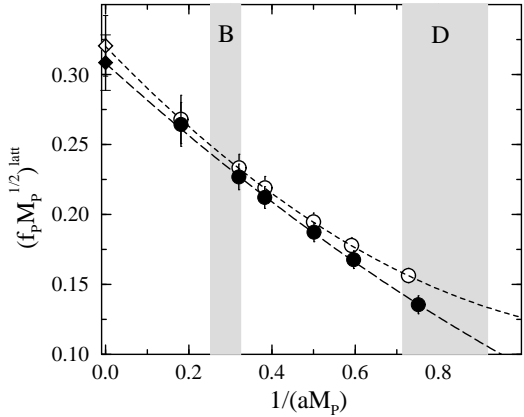}
    \vspace{-10mm}
    \caption{$1/M_P$ dependence of decay constant with the
      $O(1/m_{Q})$ action (open circles) and the
      $O(1/m_{Q}^{2})$ action (filled circles). }
    \label{fig:fsqrtM}
  \end{center}
\end{figure}

We show our results of $f_P\sqrt{m_P}$ in Figure
\ref{fig:fsqrtM}.
We find that the size of $O(1/m_P^2)$ correction is as small
as about 3 \% around the $B$ meson region and about 15 \% around 
the $D$ meson region.

\begin{figure}[tb]
  \begin{center}
    \epsfxsize=7.0cm \epsffile{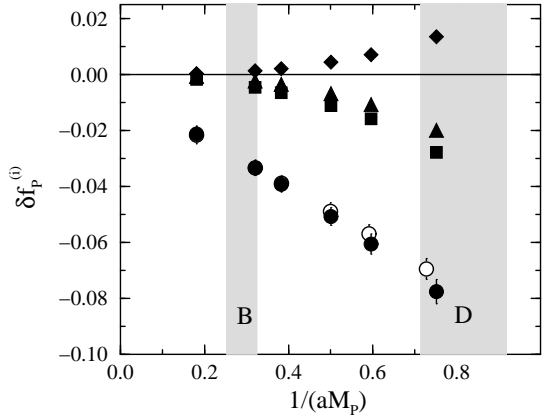}
    \vspace{-10mm}
    \caption{Corrections to the decay constant from each
      term of the current rotation.
      Circles represent the correction of $1/m_{Q}$ term 
      (\ref{eq:R(2)}) with $O(1/m_{Q})$ (open circles) and 
      $O(1/m_{Q})^{2}$ (filled circles) actions.
      Squares, diamonds and triangles are corresponding to
      the corrections from $1/m_{Q}^{2}$ terms
      (\ref{eq:R(3)}), (\ref{eq:R(4)}), and (\ref{eq:R(5)})
      respectively. } 
    \label{fig:correction}
  \end{center}
\end{figure}

In order to see how the correction terms in the rotation of 
operator (\ref{eq:R(2)})--(\ref{eq:R(5)}) changes the
result, we show the contributions from each correction term
to $f_P \sqrt{M_P}$ in Figure \ref{fig:correction}. 
We find that $O(1/m_Q)$ corrections are rather large. 
On the other hand, each of the three $O(1/m_Q^2)$ correction is
about 2 \% around the $B$ meson region.
There is a cancellation among the three corrections, and the 
total effect is of 3\%.
Although the effect of the higher order corrections is
naively expected to be very small, there is no guarantee
whether this cancellation takes place at higher order.
We, therefore, estimate an upper bound for the $O(1/m_P^3)$
error to be of 6\% at $B$ meson region.  

We study $1/m_P$ dependence of $f_P \sqrt{m_P}$ by fitting
the data with the following form
\begin{eqnarray}
  f_P \sqrt{m_P} & = & 
  (f_P \sqrt{m_P})^{\infty}
  ( 1 + \frac{c_1}{m_P} + \frac{c_2}{m_P^2} + \cdots ),
\nonumber
\end{eqnarray}
for which we obtain
\begin{eqnarray}
  (f_P \sqrt{m_P})^{\infty} & = & 0.308(20), \nonumber \\
  c_1             & = & -0.87(11) \nonumber, \\
  c_2             & = & 0.17(11) \nonumber. 
\end{eqnarray}
with the entire $O(1/m_Q^2)$ calculation.
In physical units $c_1=-1.49(19)GeV$,$c_2=(0.71(23) {\rm GeV})^2$,

To summarize, our analysis of $f_B$ shows that $O(1/m_Q^2)$ 
relativistic correction is of about 3 \% and $1/m_P$
expansion from the static limit has a good behavior\cite{Hiroshima}. 
In order to obtain $f_B$ with higher precision, one has to 
control other systematic errors such as perturbative and
discretization errors. 
We have not included one-loop correction for the
renormalization constant, for which the calculation is
underway. NRQCD group recently calculated the full one-loop
renormalization factor for the decay constant with nonrelativistic
heavy quark of $O(1/m_Q)$ and clover light quark\cite{Junko}.
They find that the effect of operator $\partial P_5$ which appear at 
one-loop level significantly reduces the decay constant.
We are also planning to carry out the simulations at higher
$\beta$ values with $O(a)$-improved Wilson light quark to
remove $O(a)$ error in near future. 

\section{$B$ meson semi-leptonic decay form factors}
\label{sec:form_factor} 
In this section we report our study of $B \rightarrow \pi l
\overline{\nu}$ semi-leptonic decay form factors.
This is the first calculation with the NRQCD action.
Earlier attempts to calculate the form factors were made by
APE\cite{APE}, UKQCD\cite{UKQCD}, Wuppertal\cite{Wuppertal}
 by extrapolating the
results in the D meson mass region obtained with the clover
action assuming heavy quark scaling law.
Direct simulation would be certainly necessary as an
alternative approach just as in the calculation of $f_B$ in
order to investigate the mass dependence and to obtain
reliable results. The status of form factors with Wilson/Clover
as well as Fermilab action is summarized in refs.\cite{Jim,Flynn}.

The semi-leptonic decay form factors $f^{+}$ and $f^{0}$ are
defined as follows
\begin{eqnarray}
  \lefteqn{ \langle\pi(k)|V_{\mu}|B(p)\rangle }\nonumber \\
    & = & \left( p+k-q \frac{m_B^2-m_{\pi}^2}{q^2}
          \right)_{\mu} f^{+}(q^2)  \nonumber \\
    &   & + q_{\mu} \frac{m_B^2-m_{\pi}^2}{q^2} f^{0}(q^2) 
\end{eqnarray}
where $q_{\mu}=p_{\mu}-k_{\mu}$ and $|B(p)\rangle$ has a normalization
\begin{equation}
  \label{eq:covariant_normalization}
  \langle B(p)|B(p^{\prime})\rangle =
  (2\pi)^3 2 E_B(p) \delta^3
  (\mbox{\boldmath $p$}-\mbox{\boldmath $p$}^{\prime}). 
\end{equation}
The virtual $W$ boson mass $q^2$ takes a value in a region
$0\le q^2 \le q_{max}^{2}$ where $q_{max}^2=(m_B-m_{\pi})^2$. 
In the rest frame of the initial $B$ meson, pion is also
almost at rest in the large $q^2$ region ($q^2\approx
q_{max}^2$), where the $W$ boson carries a large fraction of 
released energy from the $B$ meson.
In the small $q^2$ region ($q^2\approx$ 0), on the other
hand, the pion is strongly kicked and has a large spatial
momentum ($\mbox{\boldmath $k$}_{\pi}\sim$ a few GeV/c).
Lattice calculation is not reliably applicable for the 
small $q^2$ region, since the discretization error of
$O(a\mbox{\boldmath $k$}_{\pi})$ becomes unacceptably large.
This leads to a fundamental restriction in the kinematical
region where lattice calculation may offer a reliable
result.

The differential decay rate is given as
\begin{eqnarray}
  \lefteqn{ \frac{d\Gamma(\overline{B}^{0} 
      \rightarrow \pi^{+}l^{-}\overline{\nu})}{dq^2} }
      \nonumber \\ 
  & = & \frac{G_F^2 |V_{ub}|^2}{192\pi^3m_B^3} 
        \lambda^{3/2}(q^2)|f^{+}(q^2)|^2
\end{eqnarray}
where
\begin{equation}
  \lambda(q^2) =
  ( m_B^2 + m_{\pi}^2 -q^2 )^2 -4 m_B^2 m_{\pi}^2. 
\end{equation}
The decay rate vanishes at $q^2=q_{max}^2$ because the phase 
space gets smaller.
It is, thus, essential to calculate $f^+(q^2)$ in a $q^2$
region where the experimental data will become available and
the systematic error does not spoil the reliability, in
order to determine $|V_{ub}|$ model independently.

Another important feature of the $B$ meson semi-leptonic
decay is the implication of the Heavy Quark Effective Theory 
(HQET)\cite{HQET}.
In the heavy quark mass limit, it is more natural to
normalize the heavy meson state as
\begin{equation}
  \langle \tilde{M}(p)|\tilde{M}(p^{\prime})\rangle =
  (2\pi)^3 2 \left( \frac{E_M(p)}{m_{M}} \right) \delta^3
  (\mbox{\boldmath $p$}-\mbox{\boldmath $p$}^{\prime})
\end{equation}
instead of the covariant normalization
(\ref{eq:covariant_normalization}). 
With this normalization, the large mass scale $m_{M}$ is
removed from the theory and one can use the heavy quark
expansion.
The amplitude may be expanded as
\begin{eqnarray}
  \lefteqn{ \langle\pi(k)|V_{\mu}|\tilde{M}(p)\rangle }
  \nonumber \\
  & = & \langle\pi(k)|V_{\mu}|M(p)\rangle / \sqrt{m_{M}} 
  \nonumber \\
  & = & X_{\mu}^{\infty}(v\cdot k) \nonumber \\
  & & \times \left( 1 + \frac{c_1(v\cdot k)}{m_M} 
       +\frac{c_2(v\cdot k)}{m_M^2} + \cdots \right)
  \label{eq:1/m-expansion}
\end{eqnarray}
where $v$ is a velocity of the heavy meson and $c_1$, $c_2$,
... are functions of $v\cdot k$.
It is worth to note that the heavy quark mass extrapolation
and interpolation have to be done with $v\cdot k$ fixed. 
In the rest frame of the heavy meson, this condition implies 
fixed pion momentum, since  $v\cdot k$ becomes
$\sqrt{m_{\pi}^2 + \mbox{\boldmath $p$}_{\pi}^2}$.
We propose to study the $1/M$ dependence of the following
quantities  
\begin{eqnarray}
  \label{eq:v4}
  V_4(k,p) & \equiv &
  \frac{\langle\pi(k)|V^4(0)|B(p)\rangle}
     {\sqrt{2E_{\pi}}\sqrt{2E_{B}}}, 
  \\
  \label{eq:vi}
  V_k(k,p) & \equiv & 
  \frac{ \langle\pi(k)| \frac{1}{k^2} \sum_i
     k^i \cdot V^i(0)|B(p)\rangle }
     {\sqrt{2E_{\pi}}\sqrt{2E_{B}}},
\end{eqnarray}
which are natural generalization of $f_P \sqrt{M_P}$ for the 
the heavy-light decay constant.
Indeed these quantities are almost raw number which one
obtains in the lattice calculation as magnitudes of
corresponding three-point functions, and then free from
other ambiguities such as the choice of mass parameter of
the heavy quark and the discretization of the spatial
momenta. 

\subsection{NRQCD action and simulation parameters}

The action we used for the semi-leptonic decay differs from
that in the $f_B$ calculation. 
\begin{eqnarray}
  \lefteqn{ S = Q^{\dagger}(t,\mbox{\boldmath $x$}) [}
  \nonumber \\
  & &
  \left( 1 - \frac{aH_0}{2n} \right)^{-n} 
  U_4
  \left( 1 - \frac{aH_0}{2n} \right)^{-n}
  Q(t+1,\mbox{\boldmath $x$})
  \nonumber \\
  & &
  - \left( 1 - a\delta H \right) Q(t,\mbox{\boldmath $x$}) ],
\end{eqnarray}
where 
\begin{eqnarray}
  H_0      & = & - \frac{\Delta^{(2)}}{2m_Q},       \\
  \delta H & = & - \frac{g}{2m_Q}
     \mbox{\boldmath $\sigma$} \cdot \mbox{\boldmath $B$}.
\end{eqnarray} 
Notation is the same as in section \ref{sec:action_fB}.
The FWT transformation operator $R$ is identical to that in 
eq. (\ref{eq:R1})
Numerical simulation has been done on the same 120 gauge
configurations as was used for the decay constant. 
For the light quarks we use the Wilson fermion at $\kappa=$ 
0.1570. 
The following six sets of parameters for the heavy quark
mass and the stabilization parameter; $( m_Q, n )$
= (5.0,1), (2.6,1), (2.1,1), (1.5,2), (1.2,2), and (0.9,2). 
$m_Q=2.6$ and $0.9$ roughly correspond to $b$- and $c$-quark
masses respectively. 

\subsection{Extraction of three-point functions}

Matrix elements are extracted from the three-point
correlation functions: 
\begin{eqnarray}
  \lefteqn{ C^{(3)}_{\mu}(
    \mbox{\boldmath $k$}, \mbox{\boldmath $p$};
    t_B,t_V,t_{\pi} ) }
  \nonumber \\
  & = & \sum_{x_f,x_s}
        e^{-i p\cdot x_B} e^{-i (k-p)\cdot x_V}
  \nonumber \\
  &   & \langle 0 |
          O_B(t_B,\mbox{\boldmath $x$}_B)
          V_{\mu}^{\dag}(t_V,\mbox{\boldmath $x$}_V)
          O^{\dag}_{\pi}(t_{\pi},0)
        | 0 \rangle
  \nonumber \\
  &   & \longrightarrow
        \frac{Z_{\pi}(k)}{ 2 E_{\pi}(k) }
        \frac{Z_B(p)}{ 2 E_B(p) }
  \nonumber \\ 
  &   & \times
        e^{ -E_{\pi}(k)(t_V-t_{\pi}) }
        e^{ -E_{B}(p)(t_B-t_V) } 
  \nonumber \\
  &   & \times
        \langle B(p)| V_{\mu}^{\dag} | \pi(k) \rangle   
  \nonumber \\
  &   & \;\;\; (\mbox{for}\;\;t_B \gg t_V \gg t_{\pi})
\end{eqnarray}
where $t_B$, $t_V$ and $t_{\pi}$ indicate the location of
the initial state $B$ meson, heavy-light current and the
final state pion respectively.
We fix $t_{\pi}=4$ and $t_V=14$, and $t_B$ is a variable.
The amplitude 
$\langle\pi(k)|V_{\mu}|B(p)\rangle
/\sqrt{2E_{\pi}}\sqrt{2E_B}$
is obtained by dividing the above expression by the
corresponding two-point functions.

\subsection{Effective masses}

In this section, we present our numerical results.

In order to see whether the contamination from the excited
state is sufficiently small, we examine the effective mass
plots of the two-point functions of pion and $B$ meson, and
three-point functions.

\begin{figure}[tb]
  \begin{center}
    \epsfxsize=7.0cm \epsffile{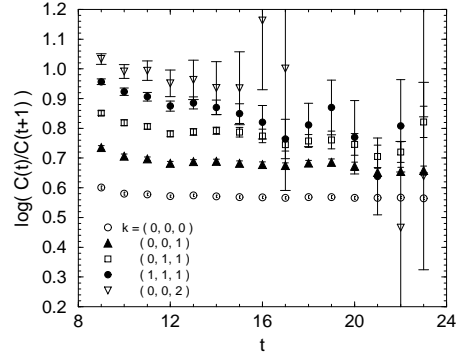}
    \vspace{-12mm}
    \caption{Effective mass plot for the pion with finite 
      spatial momenta.} 
    \label{fig:efp_pi}
  \end{center}
\end{figure}

Figure \ref{fig:efp_pi} shows the effective mass plot for
the pion with finite spatial momenta up to
 $|a\mbox{\boldmath $k$}|=2(2\pi /16)$. 
Correlation functions seem to reach to the ground state
beyond $t=14$, except for $a\mbox{\boldmath $k$}=(0,0,2)$
where statistical error becomes too large to extract reliable 
ground state energy.
We, therefore, use momentum values up to 
$|a\mbox{\boldmath $k$}|=\sqrt{3}(2\pi /16)$ in the
following analysis.
The maximum momentum value corresponds to $\sim$ 1 GeV/c in
the physical unit.

\begin{figure}[tb]
  \begin{center}
    \epsfysize=7.0cm \epsffile{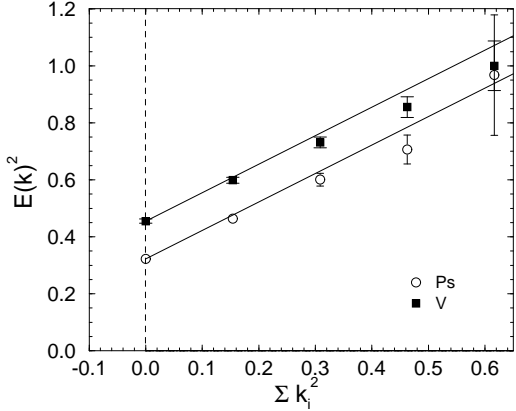}
    \vspace{-10mm}
    \caption{Dispersion relation for the $\pi$ and $\rho$
      mesons at $\kappa=0.1570$. }
  \label{fig:light_dr}
  \end{center}
\end{figure}

The momentum dependence of the pion energy should obey the
dispersion relation 
\begin{equation}
  E_{\pi}(\mbox{\boldmath $k$})^2
  = m_{\pi}^2 + \mbox{\boldmath $k$}^2
  \label{eq:dr_cont}
\end{equation}
in the continuum limit.
Then the deviation of the lattice dispersion relation from
the continuum one is a good indicator of the discretization
error. 
In figure \ref{fig:light_dr} we plot the dispersion relation 
for $\pi$ and $\rho$ mesons measured in our simulations.
Solid lines represent the above expression
(\ref{eq:dr_cont}), which does not fit the measured point,
showing the effect of the $O(a)$ discretization error.

\begin{figure}[tb]
  \begin{center}
    \epsfxsize=7.0cm \epsffile{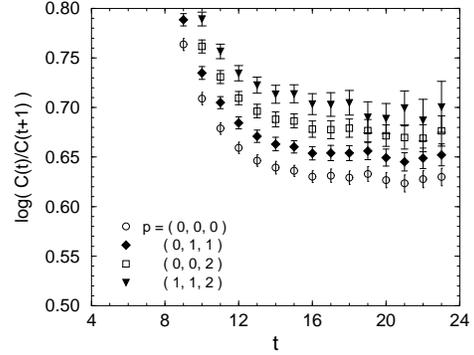}
    \vspace{-10mm}
    \caption{The effective plots for the heavy-light
      two-point function at $am_Q=2.6$ and $\kappa=0.1570$.}
    \label{fig:efp_B_L}
  \end{center}
\end{figure}

A similar plot for the $B$ meson is found in Figure
\ref{fig:efp_B_L}, where the maximum momentum is 
$|a\mbox{\boldmath $k$}|=\sqrt{3}(2\pi /16)$.
The fitting interval is chosen to be 16-24 where the
contamination of excited state is negligible for each
momentum values.

\begin{figure}[tb]
  \begin{center}
    \epsfxsize=7.0cm \epsffile{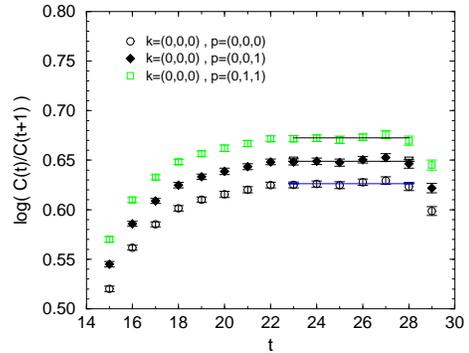}
    \vspace{-10mm}
    \caption{Effective mass plot for the three point
      function with the fourth component of the vector
      current $V_4$.} 
    \label{fig:effV4}
  \end{center}
\end{figure}

\begin{figure}[tb]
  \begin{center}
    \epsfxsize=7.0cm \epsffile{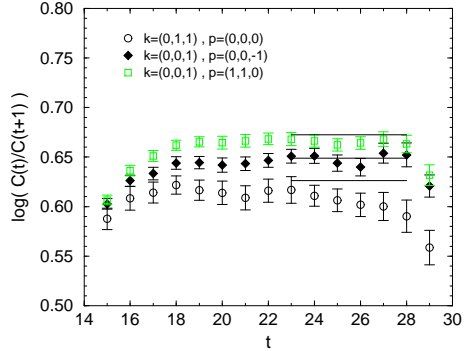}
    \vspace{-10mm}
    \caption{Effective mass plot for the three point
      function with the spatial component of the vector
      current $V_i$.} 
    \label{fig:effVi}
  \end{center}
\end{figure}

Figures \ref{fig:effV4} and \ref{fig:effVi} show the
effective mass plots for three point functions with temporal
and spatial currents respectively. 
The location of the interpolating field of the $B$ meson is 
varied to define the effective mass, then the energy should
coincide with that obtained from the two-point function if
the ground state is sufficiently isolated.
We observe clear plateau in a wide range for $t$ of 23-28 and
the mass values consistent with that from the two-point
functions, which are shown as solid lines in the figures.

\subsection{$1/m$ dependence}

\begin{figure}[tb]
  \begin{center}
    \epsfxsize=7.0cm \epsffile{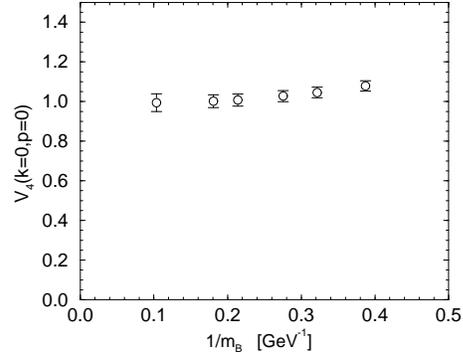}
    \vspace{-10mm}
    \caption{$1/m$ dependence of the matrix element $V_{4}$
     in eq.(\ref{eq:v4}) at zero recoil.}
    \label{fig:V4}
  \end{center}
\end{figure}

\begin{figure}[tb]
  \begin{center}
    \epsfxsize=7.0cm \epsffile{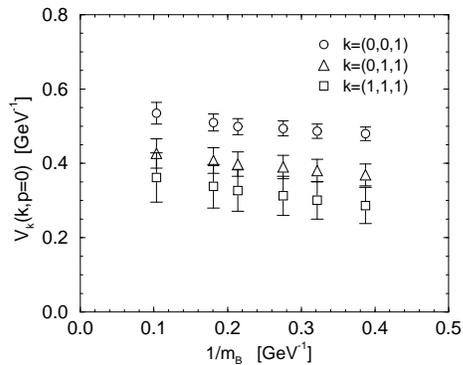}
    \vspace{-10mm}
    \caption{1/M dependence of the matrix element $V_k$
     in eq. (\ref{eq:vi}).
      Circles, diamonds and squares correspond to pion
      momenta k=(0,0,1),(0,1,1) and (1,1,1) respectively.}
    \label{fig:Vki}
  \end{center}
\end{figure}

We are now confident that the ground state is reliably
extracted for both two-point and three-point functions, and
present the results for the matrix elements.
Since $1/m$ dependence of the matrix elements is the
main issue in this study, we plot $V_4$ and $V_k$, defined in 
(\ref{eq:v4}) and (\ref{eq:vi}) respectively, which obeys
the simple heavy mass scaling law (\ref{eq:1/m-expansion}),
in Figures \ref{fig:V4} and \ref{fig:Vki}. 
We observe that $1/m$ dependence of matrix elements is
rather small in contrast to the large $1/m$ correction for
the heavy-light decay constant $f_{P}\sqrt{m_{P}}$.
Although intuitive interpretation of this result is
difficult, the smallness of the $1/m$ correction is a good
news to obtain the form factor with high precision, because
the error in setting the $b$-quark mass does not affect the
prediction. 
And also this behavior is consistent with the previous
works\cite{APE,UKQCD,Jim,Wuppertal}, in which the heavy quark mass is much 
smaller than ours.

\begin{figure}[tb]
  \begin{center}
    \epsfxsize=7.0cm \epsffile{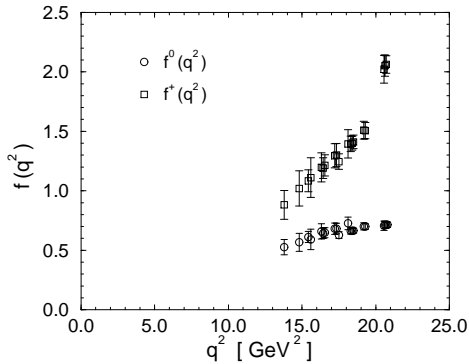}
    \vspace{-10mm}
    \caption{$q^2$ dependences of the form factors $f^0$, $f^+$.}
    \label{fig:ff0201}
  \end{center}
\end{figure}

The $q^2$ dependence of the form factors $f^+$ and $f^0$ is
shown in Figure \ref{fig:ff0201}.
The heavy quark mass is roughly corresponding to the
$b$-quark. 
As we discussed previously the accessible $q^2$ region is
rather restricted.
It is, however, interesting that already at this stage one
is able to see the momentum dependence which could really be
tested by looking at the momentum spectrum data in the
future $B$ factories. 

\section{Discussion}
\label{sec:discussion} 

We have studied $1/m_Q$ dependence of the heavy-light decay
constant and semi-leptonic decay form factors with NRQCD
action. 
We find that the error of truncating higher order
relativistic correction term is as small as 6\% for the
decay constant. 
We also find that the semi-leptonic decay form factor for 
$B \rightarrow \pi l\bar{\nu}$ has very small $1/m_Q$
dependence, which is consistent with the previous results
in the Wilson/Clover approach.  

To obtain the physical result for extracting $V_{ub}$
matrix elements, chiral limit for the light quark must be
taken and calculation of the renormalization constant $Z_V$
at one-loop is required, which is now underway.

One of the largest problem is that so far, due to low
statistics and the discretization errors of
$O((a\mbox{\boldmath $k$})^2)$, lattice calculation works
only for rather small recoil region, where the statistics of
the experimental data is not high due to the phase space
suppression.
We are planning to carry out simulations with much higher
statistics, with larger $\beta$ and with improved light quarks 
so that we can push up the accessible momentum region.
It is also true that more data from CLEO, or future
$B$ factories is required for relatively small recoil region
$|\mbox{\boldmath $k$}_{\pi}| \sim$ 1 GeV, where the lattice
calculation is most reliable. 

\section{Acknowledgment}

Numerical calculations have been done on Paragon XP/S at
INSAM (Institute for Numerical Simulations and Applied
Mathematics) in Hiroshima University.
We are grateful to S. Hioki for allowing us to use his
program to generate gauge configurations.
We would like to thank J. Shigemitsu, C.T.H. Davies,
J. Sloan and the members of JLQCD collaboration for 
useful discussions.
H.M. would like to thank the Japan Society for the Promotion
of Science for Young Scientists for a research fellowship.
S.H. is supported by Ministry of Education, Science and
Culture under grant number 09740226.
 and the members of JLQCD collaboration for 
useful discussions.
H.M. would like to thank the Japan Society for the Promotion
of Science for Young Scientists for a research fellowship.
S.H. is supported by Ministry of Education, Science and
Culture under grant number 09740226.

\end{document}